\documentclass[12pt]{article}
\usepackage{latexsym,epsfig,graphics}
\newcommand{\be}{\begin{equation}}
\newcommand{\ee}{\end{equation}}
\newcommand{\bq}{\begin{eqnarray}}
\newcommand{\eq}{\end{eqnarray}}
\newcommand{\sigta}{\sigma,\tau}
\newcommand{\intst}{\int_{0}^{p^{+}} d\sigma \int d\tau}
\begin{document}
\begin{titlepage}
\today          \hfill 
\begin{center}
\hfill    LBNL-152E \\

\vskip .5in

{\large \bf Planar Graphs On The World Sheet: The Hamiltonian
Approach}
\footnote{This work was supported 
 by the Director, Office of Science,
 Office of Basic Energy Sciences, 
 of the U.S. Department of Energy under Contract 
DE-AC02-05CH11231.}
\vskip .50in

\vskip .5in
Korkut Bardakci

{\em Department of Physics\\
University of California at Berkeley\\
   and\\
 Theoretical Physics Group\\
    Lawrence Berkeley National Laboratory\\
      University of California\\
    Berkeley, California 94720}
\end{center}

\vskip .5in

\begin{abstract}

The present work continues the program
 of summing planar Feynman graphs on the world sheet.
Although it is based on the same classical action introduced in the
earlier work, there are two important new features: 
Instead of the  path integral used in the earlier
work, the model is quantized
using the canonical algebra and the Hamiltonian picture. The new
approach has an important advantage over the old one: The ultraviolet
divergence that plagued the earlier work is absent.  Using
a family of projection operators, we are able to give an exact
 representation on the world sheet of the planar graphs of
 both the $\phi^{3}$ theory,
on which most of the previous work was based, and also of the
$\phi^{4}$ theory. We then apply the mean field approximation to determine
the structure of the ground state. In agreement with the earlier work,
 we find that the graphs of $\phi^{3}$ theory form a
dense network (condensate) on the world sheet. In the case of the 
 $\phi^{4}$ theory, graphs condense for the unphysical
(attractive) sign of the coupling, whereas there is no condensation for
the physical (repulsive) sign.

\end{abstract}
\end{titlepage}

\newpage
\renewcommand{\thepage}{\arabic{page}}
\setcounter{page}{1}
\noindent{\bf  1. Introduction}
\vskip 9pt

Several years ago, the present author and Charles Thorn initiated a
program for summing planar graphs of a given field theory [1,2].
 The model most
intensively studied so far is the $\phi^{3}$ theory, although there
have been extensions to more physical theories as well [3]. The starting
point of this program is the observation, due to G.'t Hooft, that planar 
graphs of $\phi^{3}$ theory, expressed in  light cone variables,
can be represented on the world sheet [4]. It was shown in [1] that this
representation can be derived from a local world sheet field theory. This
reformulation opened the way for the application of various field theory
techniques to the summation of planar graphs.

The present article can be thought of as a follow up to an earlier
work on the same subject [5], which we review in section 2.
 Although there is quite a bit of overlap
between the present work and [5], there are also significant differences.
 The part of the earlier work that forms the starting point of this paper
    is the path integral, based on a 
suitable action, which automatically sums the planar
graphs of massless $\phi^{3}$ model. Here, we also start with same
world sheet fields and the same action as in [5] (eq.(7)). 
 However, as it stands, this path integral 
 has two defects: The factor of
$1/(2 p^{+})$ in the propagator (eq.(1)) is missing and,  more
 seriously, it suffers from both  infrared and 
ultraviolet divergences. In [5], the infrared divergence was regulated
by discretizing one of the coordinates of the world sheet. In section 2,
we point out that this discretization is the same as compactifying the
light cone coordinate $x^{-}$ on a circle. This type of compactification
was first introduced in connection with the M theory [6,7]; it can be
viewed as an infinite boost of more standard compactification of a
spacelike direction. We take the point of view that this is not really a
serious problem, since, after all, the compactified model is  interesting 
and perfectly consistent on its own right.
 How to decompactify it is an
interesting question which will be left for future research.

The ultraviolet divergence has a more complicated origin. As explained
in section 2, it is caused by the integration over an auxilliary field
on the world sheet.  
 The solution to this problem 
proposed in [5] was to introduce a Gaussian convergence factor (eq.(8)).
This cures the original divergence, but unfortunately, it introduces
a new ultraviolet divergence, which has to be regulated by a cutoff.
The existence of this cutoff, which is really an ad hoc modification
of the theory,
 has been  a stumbling 
block to the further development of the program, since it is not clear
how to get rid of it. The standard idea of renormalization does not
seem to be applicable here. The main result of this article is to show that
when the model is quantized correctly, the original divergence 
 is absent. Therefore, there is no need for the Gaussian
convergence factor, and consequently, the ultraviolet divergence which
resulted from it is also avoided. This is the essential 
improvement introduced here over the earlier work.

In view of the problem discussed above, to what extent can we trust
the action given by (7)?
At the end of section 2, we argue that we can trust it as 
a classical action: it reproduces the correct equations of motion,
eqs.(2) and (3), for the world sheet fields.  Quantizing it by
 the path integral based on this action
  is problematic since, as we have already pointed out,
 it results in a divergent answer. 

To overcome this difficulty, we take a different, and we believe, a 
better founded
 route to quantization. Following the well known prescription of
elementary quantum mechanics,
we read off the canonical variables and the Hamiltonian from the
classical theory, and then impose the standard canonical commutation
relations. This defines the dynamics in the Heisenberg picture. We
argue in sections 3 and 4 that the model consists of a finite
number of degrees of freedom, and therefore it cannot have any
ultraviolet divergence, whose existence requires an infinite number of
degrees of freedom. As a check, the ground state energy, computed in the
mean field approximation in section 6, comes out finite.

We will carry out the program described above in two steps. First,  
the non-interacting model is treated  in section 3. In this 
simple case, the quantized Hamiltonian is easy to construct; however, 
to express it in a compact form, we found it necessary to
introduce a family of projection operators. 
 These operators and their
close relatives later play an essential role in the incorporation of the
mising $1/(2 p^{+})$ factor and in the generalization to the
$\phi^{4}$ model. The idea behind these projection operators was already
present in reference [5]; but there they were introduced in the context
of the mean field approximation. Here they are given a precise formulation
independent of any approximation scheme.

In the second step of the program, carried out in section 4, the
interaction term is taken into account. Rather than try to do this
directly in the Heisenberg picture,  it is more convenient to
take a detour and do the quantization in the interaction representation.
  The advantage of this approach is that
in between the interaction times, fields propagate freely, and the free
Hamiltonian and its quantization developed in the previous section can be
taken over without any change. Also, an extra term induced by the
interaction can be computed with relative ease.

 So far, we have been working in the Heisenberg picture based on the
Hamiltonian; but once we have
the full Hamiltonian, we can reconstruct the corresponding action, (25),
through the standard Legendre transformation. Since the action depends
on both the coordinates and their conjugate momenta, 
  to compare it to a path integral based
on an action like eq.(4), which depends  only on the coordinates
(fields), one has to carry out the integration over the momenta.
In most practical cases, the dependence on the momenta is quadratic, so
the integration can be explicitly done. In contrast, here the
momentum dependence is more complicated, and doing the momentum integration
is really not feasible. Therefore, we conclude that the action of
equation (4) makes sense only classically and it cannot be used directly
in the path integral. One has either to forego the path integral entirely
and instead work with the Hamiltonian, as we are doing here, or try to
work with the phase space version of the path integral. This point is
discussed more fully at the end of section 4.

In section 5, the missing $1/(2 p^{+})$ factor is incorporated in the
definition of the interaction vertex. This is done by means of a
projection operator, closely related to the one already used in
the construction of the free Hamiltonian. With this missing step
finally in place, we have an exact reformulation of the planar
$\phi^{3}$ theory in terms of a Hamiltonian on the world sheet.

Needless to say, although this Hamiltonian is exact, it is also quite
complicated, and one needs a manageable approximation scheme. In section
6, we introduce the mean field approximation scheme, which was already
used extensively in the past work [2,5]. This approximation scheme is
most conveniently formulated in terms of the composite field $\rho$,
defined by eq.(5): It simply amounts to replacing $\rho$ by its
ground state expectation value $\rho_{0}$. The ground state energy is
then computed as a function
of $\rho_{0}$, and minimizing it determines the value of $\rho_{0}$.
This is pretty much along the lines of the standard effective potential
calculation of the ground state [8].

In section 6, we carry out this calculation and find a non-zero value
for $\rho_{0}$, which implies the following interesting structure for
the world sheet. The world sheet
consists of two parts: The bulk, where the field $\phi$ propagates
freely, and the boundaries, marked by solid lines in Fig.1, where 
Dirichlet boundary conditions are imposed (eq.(3)). The field
$\rho$  measures the density of the solid lines (boundaries) on the
world sheet.
 A non-zero ground state
expectation value $\rho_{0}$ corresponds to a finite density of solid
lines, which we identify with a new phase of the model, and call it
the condensate phase. In this phase,
the Feynman graphs have condensed to form a dense network on the world
sheet, and the dominant contribution to the ground state comes from
the higher order graphs.
 In contrast, if $\rho_{0}$
 is zero, the density of the solid lines tends to zero in the limit
of large radius of compactification. In this phase, which we call the
perturbative phase, the main contribution to the ground state comes from
the lower order graphs. The calculation in section 6 shows that the
condensate phase has lower ground state energy
 than the perturbative phase. At the end of the section, we compare
the cutoff independent ground state energy with the cutoff dependent
result of [5], and point out both the common features and the differences.

At this point, a natural question to ask is whether  the condensation of the
Feynman graphs  leads to the formation of a string. The geometric
intuition indeed suggests that a dense network of Feynman graphs on the
world sheet should be identified with some sort of a string; in fact,
this was the picture that motivated the original work on this subject [9,10].
This picture should be confirmed or disproved by determining the spectrum
of the model and comparing it to the linearly rising trajectories of
 string theory. We hope to return to
this problem in the future.

In the next section, section 7, we apply the machinery developed in the
previous sections to the $\phi^{4}$ theory. First, we present
an exact formulation of the model on the world sheet, and then we apply
the mean field method to determine its ground state. For the physical
sign of the coupling constant, corresponding to a stable theory, nothing
interesting happens: It is the perturbative phase that is energetically
favored. On the other hand, for the ``wrong'' sign of the coupling
constant, corresponding to an unstable model, the condensate phase
has the lower ground state energy. We discuss a possible physical reason
for this correlation between instability and condensate formation.
Finally, in section 8, we summarize our conclusions and suggest some
 directions for future research.

\vskip 9pt
\noindent{\bf 2. The World Sheet Action}
\vskip 9pt

In this section, we briefly review the the world sheet action, developed in
references [1,2,5], for the planar graphs of
 $\phi^{3}$ theory in $D+2$ dimensions.
 Starting with the world sheet parametrized by the light cone variables
$$
\tau= x^{+}=(x^{0}+x^{1})/\sqrt{2},\; \sigma=p^{+}=(p^{0}+p^{1})/\sqrt{2},
$$
a general planar Feynman graph can be represented by a bunch of horizontal
 solid lines (Fig.1).
\begin{figure}[t]
\centerline{\epsfig{file=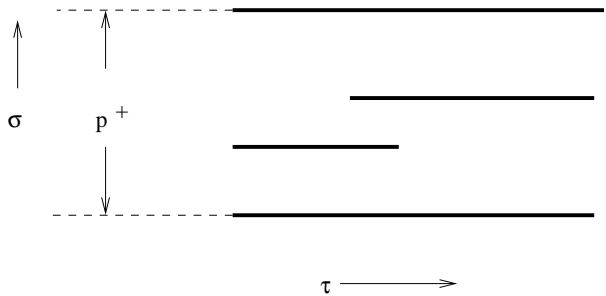,width=8cm}}
\caption{A Typical Graph}
\end{figure}
 The n'th solid line carries a transverse momentum
${\bf q}_{n}$, and two adjacent lines n and n+1, represent the light
cone propagator
\be
\Delta(p)=\frac{\theta(\tau)}{2 p^{+}}\exp\left(-i\tau\,\frac{{\bf p}^{2}
+ m^{2}}{2 p^{+}}\right),
\ee
where ${\bf p}_{n}={\bf q}_{n}-{\bf q}_{n+1}$. The interaction
 takes place at
the beginning and the end of each line, where a factor of g, the coupling
 constant, is inserted [1,4].

We remind the reader that the planar model we are studying
 is the large $N$ limit of  a field theory with the interaction
$$
\int d^{D+2} x\,g\, Tr(\phi^{3}(x)),
$$
where the field $\phi$ is an $N \times N$ Hermitian matrix. As 't Hooft 
showed [5], in this limit, each planar graph appears once and once only.
There are no local symmetry factors, such as the well known factor of $1/2$
associated with the one loop contribution to the propagator of an 
ordinary $\phi^{3}$ theory, where $\phi$ is not a matrix. There
 may be, however, global
symmetry factors associated with special configurations of momenta of
the external lines. In this paper, we are only considering generic
graphs with no global symmetries, and
 we are focusing exclusively on the
dynamics in the bulk, represented by the Hamiltonian.

In the case $m=0$,
the light cone Feynman rules sketched above can be reproduced by a field
theory that lives on the world sheet. To keep things simple, we will
study the massless model exclusively in this work, although using the tools
we are going to develop, one can easily introduce a finite mass.
 Here, we briefly describe the equivalent field theory, and
refer to [5] for the detailed derivations. The transverse momenta ${\bf q}$,
originally defined only on the solid lines, can be promoted to local
fields ${\bf q}(\sigta)$ over the whole world sheet. The solid lines form
the world sheet boundaries, and ${\bf q}$ satisfies the equation
\be
\partial^{2}_{\sigma}{\bf q}(\sigta)=0,
\ee
in the bulk, and the Dirichlet condition
\be
\partial_{\tau}{\bf q}(\sigta)=0,
\ee
on the boundaries. With the help of a Lagrange multiplier field
${\bf y}(\sigta)$, both the equations of motion and the boundary
conditions are incorporated in the following action [5]:
\be
S_{q}=\intst\left(-\frac{1}{2} {\bf q}'^{2}+ \rho\,{\bf y}\cdot
\dot{{\bf q}}\right),
\ee
where a dot represents derivative with respect to $\tau$ and a prime the
derivative with respect to $\sigma$. The $\sigma$ coordinate is compactified 
by imposing periodic boundary conditions at $\sigma=0$ and $\sigma=p^{+}$,
where $p^{+}$ is the total $+$ component of the momentum entering the
graph. The field $\rho(\sigta)$ is a
delta function on the boundaries and vanishes in the bulk: It is inserted
to ensure that the Dirichlet condition (3) is imposed only on the
boundaries.

In the functional integral, one has to integrate not only over ${\bf q}$
and ${\bf y}$, but also over the locations and the lengths of the solid
lines. This is best accomplished by introducing a two component fermion
field $\psi_{i}(\sigta)$, $i=1,2$, and its adjoint $\bar{\psi}_{i}$, and
setting
\be
\rho=\frac{1}{2} \bar{\psi}(1-\sigma_{3})\psi.
\ee
The action for the fermions is \footnote{ Here we are using the letter
$\sigma$ for both the world sheet coordinate and also for Pauli
matrices to be sandwiched between $\bar{\psi}$ and ${\psi}$. Hopefully,
there should be no confusion about the dual use of this letter.}
\be
S_{f}=\intst (i \bar{\psi} \dot{\psi} -g\,\bar{\psi}\sigma_{1}\psi).
\ee

So far, we have been treating both $\sigma$ and $\tau$ as continuous
 variables.
However, as explained in the introduction, to avoid infrared divergences, we
are going to compactify the lightcone coordinate $x^{-}$ at a radius $R$.
This is equivalent to discretizing the coordinate $\sigma$ into segments of 
length $a$, where $a= 2 \pi/R$. The discretization also makes it easy to
visualize what is going on on the world sheet. As pictured in Fig.2, the
world sheet consists of horizontal dotted and solid lines, spaced at
distance $a$ apart.
\begin{figure}[t]
\centerline{\epsfig{file=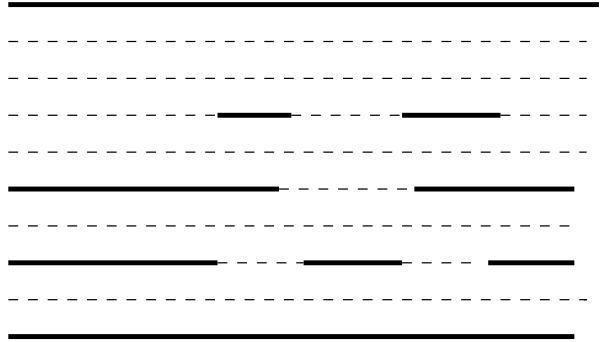,width=8cm}}
\caption{Solid And Dotted Lines}
\end{figure}
 The boundaries are marked by the solid lines,
associated with the $i=2$ component of the fermion, and the bulk is filled
with the dotted lines, associated with the $i=1$ component. The first term
in eq.(6) represents the free propagation of the fermion, tracing a solid
or dotted line, and the second term, which converts a dotted line into
a solid one or vice versa, represents the interaction. We note that there are
some extra unwanted states in the fermionic Hilbert space; they can be
eliminated by truncating it to states singly occupied
at each value of $\sigma$. 
This can be done consistently since  fermion number is
locally  conserved. 
 Integrating over the
fermion field is then the same as summing over the location and the length
of the boundaries.

 The integrals over $\sigma$ in eqs.(4,6) are in reality
finite sums with $N=p^{+}/a$ terms:
$$
\int d\sigma \leftrightarrow \sum_{\sigma}.
$$
 In order not to complicate the
notation, from time to time, we will still write an integral over $\sigma$,
although in reality it is a discrete sum. Wherever possible, we will
simplify the algebra by taking $R$ to be large, but always finite.
For example, when integrating over ${\bf q}$ ((9,10)), we used the large
$R$ (small $a$) approximation. This is not essential, but it avoids
unnecessary complications. 
 In contrast to $\sigma$,
 the time coordinate $\tau$ will remain continuous.

It is now natural to identify
\be
S=S_{q} +S_{f} 
\ee
as the candidate for the total action. There are, however, two problems
with this choice:\\
a) The exponential part of the propagator in eq.(1) is correctly
reproduced, but the prefactor
$1/(2 p^{+})$  is missing. This will be corrected in 
section 5.\\
b) A more serious drawback is that, as it stands, the functional
integral over ${\bf y}$ is divergent. This is because ${\bf y}$ lives
effectively only on the solid lines:
 The integrand is
independent of ${\bf y}$ in the bulk (on the dotted lines). Consequently,
the integrations over those ${\bf y}$ which live in the bulk are divergent. 

 The solution to this
problem, proposed in reference [5], was to introduce an additional term
given by
$$
S_{g.f}=\intst \left(-\frac{1}{2} \alpha^{2} \bar{\rho}\,{\bf y}^{2}\right),
$$
where $\alpha$ is a parameter and $\bar{\rho}$ is defined by
\be
\bar{\rho}=\frac{1}{2}\,\bar{\psi}(1+\sigma_{3})\psi,
\ee
and it is complimentary to $\rho$: It vanishes on the solid lines and provides
a Gaussian cutoff for the functional integral on the dotted lines. The
divergence mentioned above is therefore cured; but unfortunately, it is
replaced by a new  divergence: The action (8) is ultraviolet divergent
and needs a cutoff. One way to see this is to notice that in the bulk,
$S_{g.f}$ introduces a mass term for the field ${\bf y}$, but there is
no corresponding kinetic energy term in the action. A field theory with
only a mass term in the action has a constant, momentum independent
propagator and therefore it is  ultraviolet divergent. In reference [5],
the model studied was an ultraviolet regulated theory based on the action
$$
S=S_{q}+S_{f}+ S_{g.f}.
$$
Apart from the drawback  of the need for an ultraviolet cutoff, the
introduction of a new parameter $\alpha$ suggests that this model may
no longer be the original $\phi^{3}$ theory but some modification of it.

In the present work, we wish to avoid introducing any 
spurious  cutoff
or modifying the original $\phi^{3}$ theory in any way. Therefore,
our starting point will simply be eq.(4), without the additional term
$S_{g.f}$. We then face the problem of a divergent functional integral
mentioned above.

The solution to this problem, as we shall soon see, is to adopt the
Hamiltonian formulation. We would like to emphasize that, independent
of the divergence problems, the Hamiltonian approach is the correct
approach. Our reasoning goes as follows:\\
a) The action of eq.(4) is correct classically. The classical equations
derived from it are precisely the bulk equations (2) and the boundary
conditions (3). The solution to these equations reproduces any arbitrary
light cone Feynman graph for fixed boundaries as a function of the
momenta flowing through these boundaries.\\
b) This takes care of the ``classical'' part of the problem. The
``quantum'' part is the integration over the momenta and the positions
of the boundaries. Keeping the boundaries fixed for the moment, there
are two options for quantization: The first option is to
 plug in the action (4) in
the exponent and integrate functionally over the fields. The fundamental
problem with this choice is that we do not know a priori what  
measure to use in the functional integral.\\
c) The second option is to take the standard route to quantization:
Read off the canonical variables from the classical action and impose
the canonical commutation relations and finally construct the Hamiltonian.
This approach, which we are going to follow,
 is the more fundamental one. Apart from possible operator ordering
ambiguities, which are not present in this case, given the classical
action, the corresponding quantum Hamiltonian
is uniquely determined. If so desired, from the Hamiltonian in operator
form, one can derive the corresponding path integral as Feynman did and
thereby deduce the integration measure.\\
d) In the free theory, the dotted and solid lines are eternal, but the
interaction converts a dotted line into a solid one and vice versa.
This is taken care of by introducing fermions on the world sheet.
(see eq.(6)).
This formalism was extensively developed in the earlier work [2,5], and it
is applicable here without any change.\\ 
e) Although we find our canonical quantization of the classical action
quite compelling, it is still desirable to demonstrate directly that
the Hamiltonian formalism correctly reproduces the light cone
perturbation series on the world sheet. We sketch such a direct
argument at the end of section 5.

\vskip 9pt

\noindent{\bf 3. The Hamiltonian For The Free Theory}

\vskip 9pt

Rather than dealing with the interacting theory in its full complexity,
it is much easier to solve the problems discussed in the last section
in the context of the perturbation expansion in powers of
$g$. The strategy is to first construct the Hamiltonian for the
non-interacting model, and then include the interaction by going over to
the interaction representation. The starting point is the action of
eq.(7), with $g=0$. Since there is no interaction, the solid and dotted
lines are eternal, and the world sheet configuration,
 being time independent, is well suited for the Hamiltonian description. 
Let us first consider a particular graph with $n$ eternal solid lines,
that is with lines with
no beginning or end, located  at $\sigma=\sigma_{a}$, where $a$ runs from $1$
to $n$. They are ordered according to increasing $\sigma$, with
$\sigma_{a}>\sigma_{b}$ iff $a>b$.
 The action $S_{q}$ can then explicitly be written as
\be
S_{q}^{(0)}=\intst\left(-\frac{1}{2} {\bf q}'^{2}\right)+\int\, d\tau
\left(\sum_{a=1}^{n} (\dot{{\bf q}}\cdot {\bf y})_{\sigma=\sigma_{a}}
\right).
\ee

For this simple case, it is easy to avoid the
divergent functional integrals mentioned earlier.
The above action depends only on ${\bf y}(\sigma_{a},\tau)$, the Lagrange
multipliers associated with the solid lines. The divergence comes about
if one introduces additional unneeded ${\bf y}$'s located on the dotted
lines and integrates over them. Therefore, if we agree to
 integrate only over these ${\bf y}$ that live on the solid lines, there is
no longer any problem. However, this action has a serious shortcoming:
It is only applicable to a particular graph with given number of solid lines
located at specified positions, and different graphs are associated
with different actions. In a free theory, 
since the solid lines are eternal, this is not a problem,
 but this form of the action is very
awkward to generalize to the interacting theory, where the
number and positions of the solid lines can change. One could try to
 overcome this difficulty by attaching a ${\bf y}$ to every line, solid or
dotted, but as we explained earlier, one 
then is faced with the problem of a divergent functional integral. We shall
show in what follows that both of these problems can be resolved in the
Hamiltonian approach.

 We take the action (9) as our starting point for the construction
of a Hamiltonian. We choose ${\bf y}(\sigma_{a})$,
$a=1,\ldots,n$, as our canonical variables, where $\sigma_{a}$ specify
the location of the solid lines. On the other hand,
 the ${\bf q}$'s are  auxilliary variables; one can eliminate them
 in favor of the ${\bf y}$'s through the equations of motion,
\be 
{\bf q}(\sigma_{a},\tau)=\frac{1}{2}\sum_{b}\left(|\sigma_{a}
-\sigma_{b}|\,\dot{{\bf y}}(\sigma_{b},\tau)\right),
\ee
where we have dropped a term proportional to $\dot{\rho}$. This is
justified for the non-interacting theory ($g=0$), since the fermion
fields and therefore $\rho$ are time independent.

The action  can now be rewritten solely in terms of ${\bf y}$'s:
\bq
S^{(0)}&=&\int d\tau L^{(0)},\nonumber\\
L^{(0)}&=&\sum_{a,b}\left(-\frac{1}{4} |\sigma_{a}
-\sigma_{b}|\, \dot{{\bf y}}(\sigma_{a},\tau)\cdot
\dot{{\bf y}}(\sigma_{b},\tau)\right),
\eq
and  the momenta canonically conjugate to
${\bf y}(\sigma_{a})$ are:
\be
{\bf \Pi}(\sigma_{a},\tau)= - {\bf q}(\sigma_{a},\tau)=
-\frac{1}{2}\sum_{b}\left(|\sigma_{a}
-\sigma_{b}|\,\dot{{\bf y}}(\sigma_{b},\tau)\right).
\ee
Conversely, $\dot{{\bf y}}$'s can be expressed in terms of the canonical
momenta by,
\be
\dot{{\bf y}}(\sigma_{a})=\frac{{\bf \Pi}(\sigma_{a})-
{\bf \Pi}(\sigma_{a+1})}{\sigma_{a+1}-\sigma_{a}}
-\frac{{\bf \Pi}(\sigma_{a-1})- {\bf \Pi}(\sigma_{a})}
{\sigma_{a}-\sigma_{a-1}}.
\ee
Using these equations, the Hamiltonian can be written in terms of
the canonical momenta:
\bq
H_{0}&=&\sum_{a}{\bf \Pi}(\sigma_{a})\cdot \dot{{\bf y}}(\sigma_{a})-
L^{(0)}\nonumber\\
&=&\frac{1}{2}\sum_{a}\frac{\left({\bf \Pi}(\sigma_{a+1})
-{\bf \Pi}(\sigma_{a})\right)^{2}}{\sigma_{a+1}-\sigma_{a}}.
\eq

The model is quantized by letting
\be
{\bf \Pi}(\sigma_{a})\rightarrow 
- i\,\frac{\partial}{\partial {\bf y}(\sigma_{a})},\; a=1,\ldots,n.
\ee
It is clear that the eigenstates of momenta diagonalize the Hamiltonian,
 and the spectrum is a continuum starting at zero. The corresponding
Hilbert space is therefore labeled by the eigenstates of the
${\bf y}$'s associated with the solid lines, or alternatively, by the
eigenstates of the conjugate momenta. The difficulty now is that each graph
with a different set of lines is associated with a different Hilbert space,
and it is then very awkward  to deal with the interaction,
which generates transitions between these
Hilbert spaces. This problem can simply be
avoided by starting with one master Hilbert
space which includes all the possible graphs. This new Hilbert space 
is again spanned by the simultaneous eigenvalues of ${\bf y}(\sigma_{i})$ at a
fixed $\tau$. But $\sigma_{i}$ are no longer restricted to the positions
 of the
solid lines, and accordingly, ${\bf y}$ is now defined over
 the whole discretized
world sheet, including both dotted and solid lines.  Although the
interaction term causes transitions between solid and dotted lines,
the Hilbert space remains the same.

What about the Hamiltonian? The Hamiltonian is  given by eq.(14),
with $\sigma_{a}$ restricted to the positions of solid lines as before.
The ${\bf \Pi}$'s associated with the dotted lines are not present
in the Hamiltonian, and the corresponding ${\bf y}$'s are constants
of motion, fixed by the initial conditions. Since, unlike in the action
formulation, there is no integration over these ${\bf y}$'s, there
is no divergence problem in the Hamiltonian formulation.

There still remains one final obstacle to surmount. Although there is now 
only one Hilbert space for all different graphs, there is a different
Hamiltonian for each graph of the free theory.
 This is because the  sum in (14)
is restricted to the positions of the solid lines, and the location
and the number of solid lines change from graph to graph. If the
interaction is turned on, there will be transitions between different
Hamiltonians, again an awkward situation. Fortunately, the world sheet
fermion field, introduced  earlier [5] and reviewed
in section 2, was invented to solve just this kind of a problem. So far, in
constructing the Hamiltonian, we did not need the fermions,
but now they become indispensable. Recall that in eq.(14), $\sigma_{a}$ and
$\sigma_{a+1}$ refer to the locations of two successive solid lines,
seperated by only the dotted lines (Fig.(3)). We note that this is also
the configuration for a free propagator; after all, the free Hamiltonian
is nothing but a collection of free propagators. Out of the fermionic
fields, we wish to construct an projection operator which automatically
 selects such
configurations. Let us now
 define, for any two
$\sigma_{i}$ and $\sigma_{j}$, with $\sigma_{j}>\sigma_{i}$,
\be
\mathcal{F}(\sigma_{i},\sigma_{j})=\rho(\sigma_{i})\left(\prod_{k=i+1}^{k=j-1}
\bar{\rho}(\sigma_{k})\right)\, \rho(\sigma_{j}),
\ee
where $\rho$ and $\bar{\rho}$ are defined by eqs.(5) and (8), and they are
located at the same $\tau$. We usually suppress the
dependence on time coordinate $\tau$,
but in the definition of the Hamiltonian, for example, in
eq.(18), it is understood that all fields are at the same
$\tau$. We recall that
$\rho$ is one on solid lines and zero on the dotted ones, whereas $\bar{\rho}$
is zero on the solid lines and one on the dotted ones, with
$$
\rho+\bar{\rho}=1.
$$
 Our notation is such that
$\sigma_{a}$, $\sigma_{b}$ etc. refer to the positions of only the solid
 lines, whereas $\sigma_{i}$, $\sigma_{j}$ etc. refer to the positions of both
the solid and the dotted lines.

From this definition, it follows that 
\be
\mathcal{F}(\sigma_{i},\sigma_{j})=1,
\ee
if and
only if two solid lines are located at $\sigma_{i}$ and $\sigma_{j}$,
seperated only  by dotted lines, as in Fig.3.
\begin{figure}[t]
\centerline{\epsfig{file=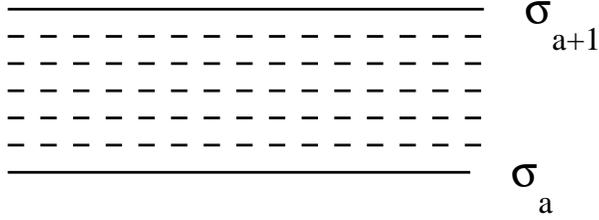,width=8cm}}
\caption{Solid Lines Seperated By Only Dotted Lines}
\end{figure}
 We recall that the Hamiltonian
(14) receives contribution only from these configurations, and
 for all other configurations, 
$\mathcal{F}(\sigma_{i},\sigma_{j})$ is zero. With the aid of this
projection operator, eq.(14) can be rewritten as
\be
H_{0}=\frac{1}{4}\sum_{i,j}\left(\frac{({\bf \Pi}(\sigma_{j})
-{\bf \Pi}(\sigma_{i}))^{2}}{|\sigma_{j}-\sigma_{i}|}\,
\mathcal{F}(\sigma_{i},\sigma_{j})\right),
\ee
where, the sum over $i$ and $j$ are unrestricted.

To the above Hamiltonian, we have to add the Hamiltonian for the fermion
field. But in the non-interacting theory, with $g=0$, and the fermionic
Hamiltonian is  zero. We can now quantize the bosonic sector as in
eq.(15) and the fermionic sector by imposing the standard anticommutation
relations. As explained in section 2, we also impose the constraint
that fermion number at each site is always one. $\rho(\sigma)$
has then two possible eigenvalues for each $\sigma$: $\rho=1$ labeling
solid lines and $\rho=0$, labeling dotted lines.  The Hilbert space,
being a superposition of these possibilities,
represents all  free graphs, with all possible combinations of
eternal solid and dotted lines. After quantization, $H_{0}$
becomes a master Hamiltonian operating in this Hilbert space, and
the problems discussed earlier, involving a multitude of Hamiltonians
and Hilbert spaces, are resolved. In the next section,
we will generalize the present treatment to the interacting theory.

\vskip 9pt

\noindent{\bf 4. The Hamiltonian For The Interacting Theory}

\vskip 9pt

It may at first appear that the Hamiltonian of the interacting
theory is the Hamiltonian of the free theory, plus the interaction
term
$$
g\sum_{\sigma}\bar{\psi}\sigma_{1}\psi.
$$
This is not, however, the end of the story; the inclusion of
interaction presents a new problem: We recall that
when   ${\bf q}$ was eliminated in favor of ${\bf y}$
through the equations of motion (10), there was a
  step that involved  integration by parts
of the time derivative in the action (4):
\be
\intst\,\rho\, {\bf y}\cdot \dot{{\bf q}}\rightarrow
- \intst(\dot{\rho}\,{\bf y}\cdot {\bf q}+ \rho\,\dot{{\bf y}}
\cdot {\bf q}).
\ee 
In deriving eq.(9), the term proportional to
 $\dot{\rho}$ on the right hand side of this equation
was dropped, since
in the free theory,  $\rho$ is time
independent. In the interacting theory, this is no longer the case.
 Notice that this term is only non-zero at the points
where transitions between the solid and dotted lines take place,
and therefore it can be thought of as part of the interaction. One 
can see this explicitly by transforming the fermion field by
$$
\psi\rightarrow \exp\left( i \frac{1-\sigma_{3}}{2} {\bf y}\cdot
{\bf q}\right) \psi,\;\;\bar{\psi}\rightarrow \bar{\psi}\,
\exp\left( - i\frac{1-\sigma_{3}}{2} {\bf y}\cdot {\bf q}\right).
$$
Under this transformation, the fermion kinetic energy  transforms 
according to
\be
i\bar{\psi}\dot{\psi}\rightarrow i\bar{\psi}
\dot{\psi}+ \dot{\rho}\, {\bf y}\cdot {\bf q},
\ee
and the unwanted
term, $\dot{\rho}\, {\bf y}\cdot {\bf q}$, cancels between  eqs.(19) and (20).
All the other terms in the action  are unchanged, except for
the interaction term
\be
g\sum_{\sigma}
\bar{\psi}\sigma_{1}\psi \rightarrow g\sum_{\sigma}
\bar{\psi}\left(\cos({\bf y}
\cdot {\bf q})\,\sigma_{1} - \sin({\bf y}\cdot {\bf q})\,\sigma_{2}\right)
\psi.
\ee

This is then the new interaction term that has to be added to the
free Hamiltonian $H_{0}$. The resulting total Hamiltonian
 is, however, difficult to handle as it stands. For
one thing, we would like to eliminate the auxilliary field ${\bf q}$
in favor of ${\bf y}$ and ${\bf \Pi}$, as we did in the case of
 the free theory. But since the equations of motion
for ${\bf q}$  are now
non-linear, this can no longer be done explicitly. Also, such a complicated
theory is difficult to quantize directly. Both of these problems can be
overcome by quantizing the model in the interaction representation.
For the convenience of the reader, in the next paragraph,
 we briefly describe the interaction representation, as it is needed for
the problem at hand.

Consider the operator $U$
  responsible for the transition from the initial states at
 $\tau=\tau_{i}$ to final states at $\tau=\tau_{f}$, and expand it
in powers of $g$. The n'th term in the perturbation expansion
of this operator can  be
written as
$$
U_{n}=\frac{g^{n}}{n!} \int d\tau_{1}\cdots \int d\tau_{n}\, T\left(
H_{I}(\tau_{1})\ldots H_{I}(\tau_{n})\right),
$$
where T stands for time ordering. The important point is that the
fields that appear in the definition of $H_{I}$ are all free fields:
For example, ${\bf y}$ and ${\bf \Pi}$
 evolve according to $H_{0}$ (eq.(18)), and satisfy the canonical
algebra (15).

This still leaves the question of what to do with ${\bf q}$.
Making  use of the interaction picture, we will show that one
can set
\be
{\bf q}= - {\bf \Pi},
\ee
in $H_{I}$,
where ${\bf \Pi}$ is the freely propagating momentum conjugate
to ${\bf y}$, introduced in section 3. We have already established this
relation for the free theory, but only on  solid lines (eq.(12)).
We are therefore entitled to use it in the interaction picture,
provided that  the ${\bf q}$ that appears in $H_{I}$ is located
on a solid  line. However, as we have argued earlier,
  the interaction takes place at the transition
point between  solid and a dotted lines, and
to complicate matters, it can easily be shown
from the equations of motion that ${\bf q}$ is discontinuous at this
point. This ambiguity is resolved by noticing that we are dealing with an
 end point contribution
coming from the partial integration with respect to $\tau$ in (19).
Consequently, the ${\bf q}$ in question is located either at the beginning
or the end of a solid line. We can therefore use (22) and rewrite the
 interaction Hamiltonian in the interaction representation as
\be
H_{I}= g\sum_{\sigma}
\bar{\psi}\left(\cos({\bf y}
\cdot {\bf \Pi})\,\sigma_{1} + \sin({\bf y}\cdot {\bf \Pi})\,\sigma_{2}\right)
\psi.
\ee

Now that the interaction is written in terms of the canonical
coordinates and momenta, it is possible to exit from the interaction
representation and go back to the Hamiltonian picture by setting
\be
H=H_{0}+ H_{I}+H_{f},
\ee
where $H_{0}$ is given by (18) and $H_{I}$ by (23), and
$$
H_{f}=\sum_{\sigma} g\,\bar{\psi}\sigma_{1}\psi.
$$
 Of course, the fields 
in the Hamiltonian picture 
 are no longer free fields. Instead, they satisfy the equations of
motion generated by the full interacting Hamiltonian.
 The main purpose of the detour
via the interaction representation was to justify the replacement (22).

 So far we have adopted the Hamiltonian approach, but
we can now easily construct the action
corresponding  to $H$, through the well known relation,
\be
S=\int d\tau\left(\sum_{\sigma}{\bf \Pi}\cdot\dot{{\bf y}}- H\right).
\ee
Since $S$ is defined over the phase space, the 
independent variables of integration
in the corresponding functional integral are both ${\bf y}$ and ${\bf \Pi}$.
Although we shall not need this action in the rest of this work, it
is of some interest to compare it with eq.(4), our starting point,
to see how it avoids the problem of the divergent functional integration.
The crucial difference is that unlike in (25), where both the coordinates
and the momenta are independent variables of integration,
 the functional integration
in (4) is over only ${\bf y}$.

 To see what is going on more clearly, we
imagine carrying out the functional integral over the fermion
field. The result is a sum over functional integrals, each associated
with a different set of solid lines (see the discussion following eq.(15)).
Consider now a typical term in this sum. The Hamiltonian $H$ depends only
on the ${\bf \Pi}$'s that are located on the solid lines; the  ${\bf \Pi}$'s
that live on the dotted lines are absent. We recall that the functional
 integral over action (4) was divergent because that action was independent
of the variables located at the dotted lines. In contrast, the above
action still depends on those momenta 
through the first term on the right. Integrating over them
gives the  equations
\be
\dot{{\bf y}}(\sigma,\tau)=0,
\ee
for only those $\sigma$ located at the dotted lines.
 As we have already seen earlier, the field ${\bf y}$ is constrained to
be $\tau$ independent on the dotted lines, and the corresponding
momenta act as Lagrange multipliers enforcing this constraint.

We would like to stress that the form of the action where both the coordinates
and the momenta appear as independent variables is more fundamental: That is
the form that comes out of the usual derivation of the path integral.
One can then pass to the coordinate version of the path integral
 by integrating over the momenta. This integration is
usually easy to do since the dependence on the momenta is quadratic.
 However, in this case, the action (25) is
not quadratic in the momenta; as we have just seen,  the dependence
on some of the momenta is in fact linear. Integration over these momenta
resulted in the constraints (26), which are missing in the formulation
based on (4).
 Alternatively, one may try
to impose these constraints  by hand, but since they change depending
on the location of the dotted lines, there does not seem to be any
practical way of doing it. We conclude that there is no practical
alternative to the phase space form of the action;
 the form of the action in terms of coordinates only
that was used  [5] resulted in an ill defined path integral.

\vskip 9pt

\noindent{\bf 5. Vertex Correction}

\vskip 9pt

We pointed out earlier that the world sheet theory we have so far does not 
reproduce the prefactor  $1/(2 p^{+})$ in the propagator (eq.(1)).
This problem was addressed in [5] in the context of the mean field
approximation. Here, using the projection operator technique,
 we will  incorporate this factor
into the Hamiltonian.

Although originally the factor  $1/(2 p^{+})$ was part of the
 propagator, it turns out that it is more convenient to associate 
 with the vertices. This is done by attaching it either to
the beginning or the end of the propagator, and then transferring to the
vertex located at that position. Consider the two basic vertices pictured in
Fig.4 a,b.
\begin{figure}[t]
\centerline{\epsfig{file=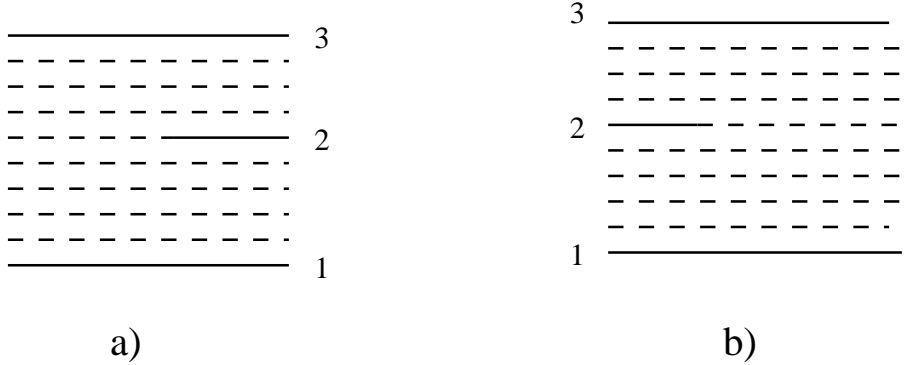,width=12cm}}
\caption{The Two $\phi^{3}$ Vertices}
\end{figure}
 In the first one,  the $+$ vertex,
 a dotted line turns into a solid one, and in the
other, the $-$ vertex, the reverse takes place. The solid lines are
labeled as 1,2 and 3, as shown in the figure, and the
propagators are labeled by the corresponding pair of indices, 12, 23 and 13
respectively.
 Let us call the extra factor to be attached to the first
vertex $V_{-}$ and the factor that goes with the second vertex $V_{+}$.
The only thing that is fixed is their product:
\be
V_{+}\,V_{-}= \frac{1}{8\, p_{12}^{+}\, p_{23}^{+}\, p_{13}^{+}},
\ee
and one is free to choose one of them arbitrarily, so long as the product 
satisfies the above relation. We make the natural symmetric choice
\footnote{This choice is different
from the one in reference [5].}
\be
V=V_{+}=V_{-}=\frac{1}{\sqrt{8\, p_{12}^{+}\, p_{23}^{+}\, p_{13}^{+}}},
\ee
which simplifies the resulting algebra.

We are now ready to combine the $V$'s with $H_{I}$
in order to supply the missing $1/(2 p^{+})$ factors. Defining
\be
\mathcal{V}_{\pm}=\bar{\psi}\,\sigma_{\pm}\,\psi\,\exp\left(\mp i {\bf y}
\cdot {\bf \Pi}\right),
\ee 
we can ,  schematically, rewrite the full interaction Hamiltonian,
incorporating the  $1/(2 p^{+})$ factors, in the following form:
\be
H_{I}\rightarrow g \left(\mathcal{V}_{+}+\mathcal{V}_{-}\right)\,
V\,\mathcal{G}
\ee

In this equation,
 $\mathcal{G}$ is a projection operator which is inserted to 
automatically choose the correct vertex configurations. By this, we mean
the following: Both vertices in Fig.4
 consist of three propagators meeting at a point, and each propagator
 consists of two solid lines seperated only by  dotted lines.
The correct vertex configuration is chosen by attaching
$\mathcal{G}$ to the propagator bounded by the solid lines 1 and 3
in Fig.4. This
operator should  be unity when the two solid lines are seperated
by only dotted lines, and zero otherwise. We already faced exactly the same
problem in deriving eq.(18), and we could use here the same projection
 operator,
 $\mathcal{F}$. It is, however, more convenient 
to define a slightly different operator:
\be
\mathcal{G}(\sigma_{i},\sigma_{j})=\left(\prod_{k=i+1}^{k=j-1}
\bar{\rho}(\sigma_{k})\right)\,\rho(\sigma_{j}).
\ee
The relation between the two projection operators is,
$$
\mathcal{F}(\sigma_{i},\sigma_{j})=\rho(\sigma_{i})\,\mathcal{G}(\sigma_{i},
\sigma_{j}).
$$
We will later see why $\mathcal{G}$ is better suited for the later
computations.

Eq.(30) can now be rewritten in a more explicit form:
\be
H_{I}= g\sum_{\sigma_{2}}\sum_{\sigma_{1}<\sigma_{2}}
\sum_{\sigma_{3}>\sigma_{2}}
\frac{\mathcal{G}(\sigma_{1},\sigma_{3})\left(\mathcal{V}_{+}(\sigma_{2})
+\mathcal{V}_{-}(\sigma_{2})\right)}
{\sqrt{8\,(\sigma_{2}-\sigma_{1})
(\sigma_{3}-\sigma_{2})(\sigma_{3}-\sigma_{1})}},
\ee
where $\sigma_{1,2,3}$ are the coordinates of the lines 1,2 and 3 in
Fig.4. Notice that the denominators in this expression never vanish:
The distance between the $\sigma$ coordinates is at least $a$, the unit
of the lattice spacing. A possible infrared divergence is therefore
avoided by the compactification of the coordinate $x^{-}$ at radius
$R=2\pi/a$ (see the discussion following eq.(6)).

To the above expression for the interaction Hamiltonian, one has to add
the free Hamiltonian $H_{0}$ (eq.(18)), to arrive at the total Hamiltonian
$$
H=H_{0}+H_{I}.
$$
We would like to emphasize that in  this final form, the Hamiltonian
is exact: Nothing has been omitted and as yet, no approximations have
been made. If so desired, one could also go over to the path integral
formulation through eq.(25). Of course, as would be expected, the
expression for $H_{I}$ is quite complicated; for example, it is 
non-local in $\sigma$. However, it is local in the time coordinate
$\tau$, which makes the Hamiltonian formulation possible. Of course,
 one cannot hope to make
progress without  some  approximation. In the next section,
we will see that a great simplification results when one applies the
mean field approximation; for example, one can determine the ground state of 
the model. Independent of any approximation, however, one thing
should be clear: We have already seen above that there is no infrared
divergence. Since 
 what we have here is a quantum mechanical system with
finite degrees of freedom, and there can also be no ultraviolet divergence.  
We therefore have a perfectly finite model. In the next section, we will
verify this in the mean field approximation.

Before closing this section, we would like to demonstrate directly that the
Hamiltonian formalism we have developed correctly reproduces the light
cone Feynman graphs on the world sheet.  Starting with the free Hamiltonian
of eq.(14), 
we will  show that it generates the  free theory
on the world sheet.
 Using eq.(12), we can make the replacement (eq.(22)),
$$
{\bf \Pi}(\sigma_{a})\rightarrow -{\bf q}(\sigma_{a}),
$$
on the solid lines, and note that,
 the time evolution operator
$$
\exp\left(- i H_{0} \,\tau\right)
$$
generates a collection of free propagators (eq.(1)),
with solid lines as boundaries,  without,
 however, the prefactor $1/(2 p^{+})$ and for $m=0$.
 But since the free theory is nothing
but the collection of free propagators, this completes the demonstration.

Next, consider the interaction, which causes transitions between solid
and dotted lines. The vertices responsible for these transitions are
represented by the term 
$$
g\sum_{\sigma}\bar{\psi}\sigma_{1}\psi
$$
in the fermionic part of the Hamiltonian. The missing prefactor of the
propagators can be incorporated into these vertices (see eq.(32)). The
use of world sheet fermions as a tool for building interaction
vertices was fully developed in references [2,5] and reviewed to some
extent in the preceding sections. We therefore conclude that 
the interacting theory is also
correctly reproduced by the Hamiltonian approach.

\vskip 9pt

\noindent{\bf 6. The Mean Field Approximation}

\vskip 9pt

The mean field approximation is a statistical approximation scheme which
has been long in use in both field theory and many body physics with
varying degree of success. In the context of the problem at hand, the
basic idea is to replace the operator $\rho$ (eq.5) by its ground state
expectation value $\rho_{0}$:
\be
\rho\rightarrow \rho_{0}=\langle \rho \rangle.
\ee
To do this systematically, it is convenient first to promote $\rho$ to
be an independent field by adding the term
\be
\Delta S=\intst\, \lambda\left(\rho-\frac{1}{2} \bar{\psi}
(1-\sigma_{3})\psi\right)
\ee
to the action, where $\lambda$ is a Lagrange multiplier field.
An alternative approach, which we will not pursue and which
gives the same results obtained here, is to bosonize
the fermions. We can
then in principle compute the ground state energy as a function of 
$\rho$  after integrating over the other fields. $\rho_{0}$ is then
 the classical field configuration that minimizes
this energy.

Originally, being a composite of the fermions, $\rho$ could only take
on the values 0 and 1; after being promoted to an independent field,
$\rho$ and hence $\rho_{0}$ can take on any value between 0 and 1.
There is also a statistical interpretation, derivable from the
 Euclidean version of the  path integral.
 Remembering that $\rho=1$ corresponds
to a solid line and $\rho=0$ to a dotted line, $\rho_{0}$ is then the
probability of finding a solid line at a given location. In general,
$\rho_{0}$ could depend on the coordinates $\sigma$ and $\tau$;
however, with the boundary conditions we have chosen (see the discussion
following eq.(4)), it is natural to assume that the ground state of
the system is translationally invariant in both directions. Consequently,
$\rho_{0}$ can be taken to be a constant. As we shall see shortly, replacing
$\rho$ by a constant  leads to an enormous simplification.

It is natural to ask how good this approximation is. If we set
$$
\rho=\rho_{0}+ \Delta\rho,
$$
what we are doing is to neglect $\Delta\rho$, the quantum fluctuation
around the classical field $\rho_{0}$.
In the previous work [2,5], it was argued that the fluctuation term is
suppressed by a factor of $1/\sqrt{D}$, where $D$ is the dimension of the
transverse space. This is at best a formal argument, since in practice,
$D$ is not necessarily large \footnote{$D=2$ for the usual space-time.}.
Here we will not try to justify the mean field method; instead, we will
 simply point out a necessary condition for its validity.
 It is clear that statistical methods can only
be successfully applied to problems with large degrees of freedom. In
the present case, the number of degrees of freedom is roughly
proportional to $N= p^{+}/a$, the total number of lines, dotted or
solid, on the world sheet. Large number of degrees of freedom means
small $a$ or equivalently, large $R$, the
 radius of compactification. It follows that, 
only in the limit of large $R$, which we have assumed earlier to simplify
the algebra,
 can we hope to get reasonable results
from the mean field method.

Next, we are going to compute the ground state energy of
the system as a function of $\rho_{0}$, and determine $\rho_{0}$ by
minimizing the energy. The crucial question is, whether in the limit of
large $N$, $\rho_{0}$ tends to zero or stays finite. It follows from its
meaning as probability that $\rho_{0}$ represents roughly the proportion
of the world sheet area covered by solid lines. A vanishing value for it
means that the dotted lines (bulk) greatly outnumbers the solid lines
(boundary). We note that this is the case for any fixed order in perturbation
expansion: In the limit of large $N$, the dotted lines dominate over
the solid lines, and therefore, we can conclude that the ground state
of the system gets contribution mainly from fixed order terms in the
perturbation expansion. In contrast, if $\rho_{0}\neq 0$, the solid
and dotted lines are in a finite ratio. This can only happen if as
$N$ becomes large, increasingly higher order perturbation terms
contribute to the ground state. This can be thought of as an unusual
phase of the model where the solid lines have condensed on the world
sheet, and $\rho_{0}$ can be identified with the corresponding order
parameter [5]. We will shortly see that, at least in the mean field
approximation, such a condensation takes place.

With these preliminaries out of the way, let us apply the mean field
method  to the the two terms, $H_{0}$ and $H_{I}$. As explained
above, all we have to do is replace $\mathcal{F}$ in eq.(18) and
$\mathcal{G}$ in eq.(32) by their expectation values. This amounts
to setting
$$
\rho\rightarrow \rho_{0},\;\;\bar{\rho}\rightarrow 1- \rho_{0},
$$
in their definition, with the result that
\be
\langle \mathcal{F}(\sigma_{i},\sigma_{j})\rangle=
\rho_{0}^{2}\,(1-\rho_{0})^{(\sigma_{j}-\sigma_{i})/a -1},\;\;
\langle \mathcal{G}(\sigma_{i},\sigma_{j})\rangle=
\rho_{0}\,(1-\rho_{0})^{(\sigma_{j}-\sigma_{i})/a -1},
\ee
and consequently, we have,
\be
H_{0}\rightarrow \frac{\rho_{0}^{2}}{4}\sum_{i,j}\left(
\frac{({\bf \Pi}(\sigma_{j})-{\bf \Pi}(\sigma_{i}))^{2}}
{|\sigma_{j}-\sigma_{i}|}\,(1-\rho_{0})^{|\sigma_{j}-\sigma_{i}|/a
-1}\right).
\ee
This expression can be simplified further, but to no great advantage.
 It is already quadratic in the momenta and
simple enough as it stands.

Next, in eq.(32),  carrying out the statistical averaging explained above,
the sums over $\sigma_{1}$ and $\sigma_{3}$, with $\sigma_{2}$ fixed,
 can be done. Defining,
\be
\sum_{\sigma_{1}<\sigma_{2}}\sum_{\sigma_{3}>\sigma_{2}}
\frac{\langle \mathcal{G}(\sigma_{1},\sigma_{3})\rangle}
{\sqrt{8 (\sigma_{3}-\sigma_{1})(\sigma_{2}-\sigma_{1})
(\sigma_{3}-\sigma_{2})}}
 = A(\rho_{0}),
\ee
we have,
\be
H_{I}\rightarrow g\,A(\rho_{0}) 
\sum_{\sigma}\left( \exp(-i{\bf y}\cdot{\bf \Pi})\,
\bar{\psi}\sigma_{+}\psi + \exp(i{\bf y}\cdot{\bf \Pi})\,
\bar{\psi}\sigma_{-}\psi\right),
\ee
where,
\be
A(\rho_{0})= \frac{\rho_{0}}{\sqrt{8\, a^{3}}}
\sum_{n_{1}=0}^{\infty}\sum_{n_{2}=0}^{\infty}\,\frac{ (1-\rho_{0})^
{n_{1}+ n_{2}+1}}{\sqrt{(n_{1}+ n_{2}+2)(n_{1}+1)(n_{2}+1)}}
\ee 

In deriving eq.(39), we have assumed that the upper limit of the
 sum over the $\sigma$'s can be extended to infinity, whereas in
reality there is a cutoff at $\sigma=p^{+}$, where $p^{+}$ is the plus
component of the total momentum entering the graph. The neglect of this
cutoff makes a difference only for small values of $\rho_{0}$.
Although it is not difficult to take care of the cutoff, the resulting
formulas become complicated. In the interests of simplicity, we will
work with the simple expressions derived above, keeping in mind the
restriction that they are not reliable for small values of $\rho_{0}$.

We can now express the Hamiltonian in terms of the
operators ${\bf y}$, ${\bf \Pi}$, $\bar{\psi}$, $\psi$ and the
parameter $\rho_{0}$. It remains to diagonalize it and minimize the
eigenvalue with respect $\rho_{0}$. It very easy to find one simple
state that partially diagonalizes the Hamiltonian; this is the state where 
all momenta are  zero:
\be
{\bf \Pi}(\sigma)|0\rangle =0.
\ee
This choice clearly diagonalizes $H_{0}$, with eigenvalue zero. This
state is also an eigenstate of the exponential operators that appear
in $H_{I}$ (eq.(23)).
 This is because these operators are dilatation operators that
scale the momenta, and the state with zero momentum is scale invariant:
\be
\exp\left(\pm\frac{i}{2}\,{\bf y}\cdot{\bf \Pi}\right)
|0\rangle= |0\rangle,\;\;\mathcal{V}_{\pm}\rightarrow\bar{\psi}
\sigma_{\pm}\psi.
\ee
 
We have just shown that $|0\rangle$ is one of the eigenstates of
 the Hamiltonian. It is also the unique state that is invariant
under translations in both $\sigma$ and $\tau$. Under our assumption
of  translation invariance of the ground state, it is therefore
 the only candidate for the ground state.

Taking into account   the contribution from $\Delta S$ (eq.(34)),
 and
$$
H_{0}|0\rangle= 0,
$$
the full Hamiltonian, acting on the ground state, can be written as a
quadratic in the fermion fields:
\be
H|0\rangle \rightarrow \sum_{\sigma}\left(g A\,
\bar{\psi}\sigma_{+}\psi+ g A\, \bar{\psi} \sigma_{-}
\psi
+\frac{\lambda_{0}}{2}\, \bar{\psi}(1-\sigma_{3})\psi
 -\lambda_{0}\, \rho_{0}\right),
\ee
where $\lambda$ has also been replaced by its constant ground state
expectation value $\lambda_{0}$.

It is now easy to complete the diagonalization. Since the 
fermions located
at different values of $\sigma$ decouple, it boils down to diagonalizing
a two by two matrix
$$
M=\frac{\lambda}{2} (1-\sigma_{3})-\lambda\,\rho_{0}
+g\,A\,\sigma_{1},
$$
and the two energy levels are the eigenstates of M, multiplied by 
$p^{+}/a$:
\be
E_{\pm}=\frac{p^{+}}{2 a}\left(\lambda_{0} (1- 2\rho_{0})\pm
\sqrt{\lambda_{0}^{2}+ 4\, g^{2}\,A^{2}}\right).
\ee

We now carry out the integral over $\lambda_{0}$, using the saddle point
method. This approximation can formally be justified by invoking
the large $p^{+}/a$ (large radius of compactification) limit. For a
more detailed treatment, see [5].
 The saddle point  is given by 
\be
\frac{\partial E_{\pm}}{\partial \lambda_{0}}=0\rightarrow
\lambda_{0}=\mp\frac{ |g| (1- 2 \rho_{0}) A(\rho_{0})}{
\sqrt{\rho_{0}- \rho_{0}^{2}}},
\ee
and the energy at the saddle point is
\be
E_{\pm}(\lambda=\lambda_{0},\rho_{0})=\pm 2 |g|\, \frac{p^{+}}{a}\,
A(\rho_{0})\,\sqrt{\rho_{0}-\rho_{0}^{2}}.
\ee

In Fig.5, the function
$$
F(x)=-\sqrt{8\,a^{3}}\, A(x)\,\sqrt{x -x^{2}},
$$
is plotted, where  $x=\rho_{0}$.
\begin{figure}[t]
\centerline{\epsfig{file=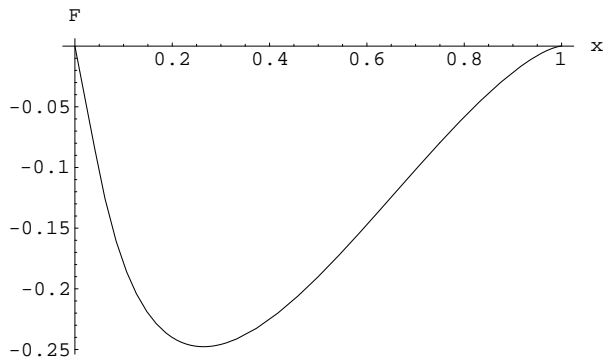,width=8cm}}
\caption{The Function F(x)}
\end{figure}
It  has a unique minimum with negative energy
 at a non-zero value of $\rho_{0}$.
As we have argued earlier, the non-vanishing of $\rho_{0}$ at the
minimum means that a finite fraction of the world sheet area is
occupied by the solid lines, or, in other words, the solid lines
have condensed. We therefore conclude that, at least in the mean
field approximation, the ground state is a condensate of the solid
lines (boundaries). On the other hand, the minimum of $E_{+}$ is 
zero, at $\rho_{0}=0$. Having a higher energy than the ground state,
this state is a false vacuum. Also, 
the vanishing of $\rho_{0}$ means that this state is in the
perturbative phase of the model.

We would like to stress that the above results are quite robust;
they do not depend on the precise functional dependence of $A$.
 For the existence of a minimum of $E_{-}$ at a
$\rho_{0}\neq 0$, all that is required is that $A(\rho_{0})$
is positive and bounded in the allowed range of $\rho_{0}$,
$$
0\leq \rho_{0} \leq 1.
$$
 Going back to eq.(37), we note that $A$ is positive because
the expectation value of $\mathcal{G}$ is positive.
 But this follows from the definition of $\mathcal{G}$:
It is the product $\rho$'s and $\bar{\rho}$'s,
which are a bunch of commuting
positive semi-definite operators.

 Altough so far we have studied the massless $\phi^{3}$ theory, so long
the mass is not too big, the results discussed above
 do not change with the introduction
of a non-zero mass. In fact, we could substantially take over the treatment
of the massive theory given in reference [5]; but in the interests of
keeping this article to a reasonable length, we will not do so.

Finally, we compare briefly the results obtained in this section to
those in [5]. 
 The expression for the ground state
energy that appeared in the earlier work is of the form
\be
E_{\pm}=E_{b}\pm E_{f}.
\ee
The first term $E_{b}$ is the contribution, in the mean field approximation,
 of the bosonic part of the action $S_{q}$ (eq.(4)), and it is positive. 
 The second term, $\pm E_{f}$,
is the contribution of the fermionic sector of the theory, which includes
both $S_{f}$ (eq.(6)), and also the prefactor $1/(2 p^{+})$ in the propagator
(see section 5). Comparing with the present work, we see that in
 eq.(45),  the counterpart
of $E_{b}$ is absent; the ground state energy comes solely from the
fermionic sector. This is the main difference between the earlier work and
 the present paper;  apart from some unimportant details,
 the term $E_{f}$ is substantially the same both here and in [5]. This explains
why we have a non-trivial ground state at $\rho_{0}\neq 0$ both here and in
the earlier work: It all comes from $E_{f}$;
 the presence or absence of a positive $E_{b}$ makes no difference.
 So we agree in
the present work with the main result of references [2,5], although, of course,
this is somewhat accidental since the expression for the ground state
energy derived here is different.

It is worthwhile noting that the ultraviolet cutoff dependence of the
energy in the earlier work was all due to $E_{b}$; $E_{f}$ is ultraviolet
finite. Where did $E_{b}$ come from? We believe that,
as explained at the end of section 4, it came from
an unjustified passage from the form of the action 
in terms of both coordinates and momenta, to a form in terms of only
coordinates. 

\vskip 9pt

\noindent{\bf 7. The $\phi^{4}$ Theory}

\vskip 9pt

The rules for the world sheet graphs of the $\phi^{4}$ theory are a modest
generalization of those for $\phi^{3}$. The propagator is still the same
(eq.(1)), and so are the boundary conditions (2) and (3). Consequently,
$S_{q}$ is unchanged, but the interaction is different. Instead of a single
vertex, there are now two vertices, located at the same $\tau$ but at 
different $\sigma$'s. There are now four possibilities, depending on whether
the solid lines are incoming or outgoing, and they are pictured in
Fig.6.
\begin{figure}[t]
\centerline{\epsfig{file=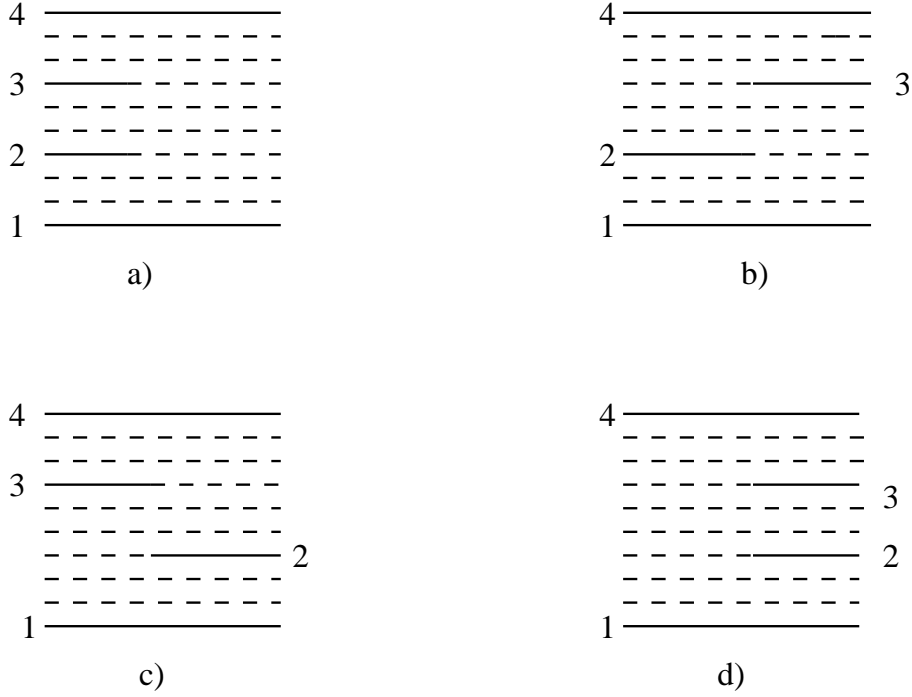,width=12cm}}
\caption{The Four $\phi^{4}$ Vertices}
\end{figure}
 The corresponding $S_{f}$, instead of eq.(6), is now given by,
\be
S_{f}=\intst\,i \bar{\psi} \dot{\psi} - g^{2} \int d\sigma
\int d\sigma' \int d\tau\,\mathcal{E}(\sigma,\sigma')\,
(\bar{\psi}\sigma_{1}\psi)_{\sigma,\tau}\,
(\bar{\psi}\sigma_{1}\psi)_{\sigma',\tau},
\ee
where,
$$
\mathcal{E}(\sigma_{i},\sigma_{j})=
\prod_{k=i+1}^{k=j-1}\bar{\rho}(\sigma_{k})
$$
is closely related to $\mathcal{F}$ and $\mathcal{G}$ introduced
earlier. It is inserted so that there are no unwanted solid lines
between the positions of the two vertices, $\sigma$ and $\sigma'$.

 We start with eq.(7), where $S_{q}$ is the same as before but
$S_{f}$ is replaced by (47). Since the free Hamiltonian depends only on
 $S_{q}$, we again end up with the same free Hamiltonian,
given by eq.(18). On the other hand, the interaction vertices are more
complicated than $\phi^{3}$ vertices. The four distinct vertices are
pictured in Figs.6,a,b,c,d. The interaction Hamiltonian is the sum
of the corresponding four terms:
\be
H_{I}= H_{I}^{a} +H_{I}^{b}+H_{I}^{c}+H_{I}^{d},
\ee
where,
\bq
H_{I}^{a}&=& g^{2} \sum_{\sigma}\frac{\mathcal{G}(\sigma_{1},\sigma_{4})
\mathcal{V}_{-}(\sigma_{2})\mathcal{V}_{-}(\sigma_{3})}{4 \sqrt{
(\sigma_{2}-\sigma_{1})(\sigma_{3}-\sigma_{2})(\sigma_{4}-\sigma_{3})
(\sigma_{4}-\sigma_{1})}},\nonumber\\
H_{I}^{b}&=& g^{2}  \sum_{\sigma}\frac{\mathcal{G}(\sigma_{1},\sigma_{4})
\mathcal{V}_{+}(\sigma_{2})\mathcal{V}_{-}(\sigma_{3})}{4 \sqrt{
(\sigma_{2}-\sigma_{1})(\sigma_{4}-\sigma_{3})(\sigma_{4}-\sigma_{2})
(\sigma_{3}-\sigma_{1})}},\nonumber\\
H_{I}^{c}&=& g^{2}  \sum_{\sigma}\frac{\mathcal{G}(\sigma_{1},\sigma_{4})
\mathcal{V}_{-}(\sigma_{2})\mathcal{V}_{+}(\sigma_{3})}{4 \sqrt{
(\sigma_{2}-\sigma_{1})(\sigma_{4}-\sigma_{3})(\sigma_{4}-\sigma_{2})
(\sigma_{3}-\sigma_{1})}},\nonumber\\
H_{I}^{d}&=& g^{2} \sum_{\sigma}\frac{\mathcal{G}(\sigma_{1},\sigma_{4})
\mathcal{V}_{-}(\sigma_{2})\mathcal{V}_{-}(\sigma_{3})}{4 \sqrt{
(\sigma_{2}-\sigma_{1})(\sigma_{3}-\sigma_{2})(\sigma_{4}-\sigma_{3})
(\sigma_{4}-\sigma_{1})}}.
\eq

In this equation, $\mathcal{V}_{\pm}$ are given by (29). The symbol
$\sum_{\sigma}$
indicates a quadruple sum over the $\sigma$'s, subject to the ordering
$$
\sigma_{4}>\sigma_{3}>\sigma_{2}>\sigma_{1}.
$$

We now search for the ground state of the system in the mean field
approximation. This involves several steps:\\
a) ${\bf \Pi}(\sigma)$ can be set equal to zero in the ground state.
The argument is the same one following (40). We can therefore make
the replacement
\be
\mathcal{V}_{\pm}\rightarrow \bar{\psi}\sigma_{\pm}\psi.
\ee
b) In the mean field approximation, $\mathcal{G}$ is replaced by its
expectation value, given by eq.(35).\\
c) The fermionic bilinears are first bosonized by adding a term
$\Delta H$ to the Hamiltonian:
\bq
\Delta H&=&\sum_{\sigma} \Bigg(\lambda\left(
\rho -\frac{1}{2}\bar{\psi}(1- \sigma_{3})
\psi \right)\nonumber\\
&+&\lambda_{1}\left(\rho_{1}-\bar{\psi}\sigma_{1}\psi\right)
+\lambda_{2}\left(\rho_{2}-\bar{\psi}\sigma_{2}\psi\right)\Bigg),
\eq
and then the $\lambda$'s and the $\rho$'s
 are replaced by their coordinate independent ground state expectation
values. In order to keep the notation simple, we will skip the
additional subscript ``0'' for the expectation value of the
corresponding field. Putting these steps together gives
\be
H|0\rangle \rightarrow \frac{p^{+}}{a}\frac{g^{2}}{a^{2}}
B(\rho)\,(\rho_{1}^{2}+\rho_{2}^{2})
+\Delta H,
\ee
where,
\bq
B(\rho)&=&\frac{\rho}{2}\sum_{n_{1},n_{2},n_{3}=0}^{\infty}\Bigg(
\frac{(1-\rho)^{n_{1}+n_{2}+n_{3}+1}}{\sqrt{(n_{1}+1)(n_{2}+1)
(n_{3}+1)(n_{1}+n_{2}+n_{3}+3)}}\nonumber\\
&+&\frac{(1-\rho)^{n_{1}+n_{2}+n_{3}+1}}{\sqrt{(n_{1}+1)(n_{1}+
+n_{2}+2)(n_{2}+n_{3}+2)(n_{3}+1)}}\Bigg).
\eq
In what follows, the detailed form of B will not be important; all that
will matter is that B is positive and bounded for $0 \leq \rho \leq 1$,
the physical range of $\rho$.
 
Repeating the steps following eq.(42), we diagonalize $\Delta H$,
a direct sum of two by two matrices. The resulting two  eigenvalues
of the ground state energy are
\be
E_{\pm}=\frac{p^{+}}{a}\left(\frac{g^{2}}{a^{2}} B(\rho)(\rho_{1}^{2}
+\rho_{2}^{2})-\frac{\lambda}{2} \pm \frac{1}{2}\sqrt{\lambda^{2}+
4(\lambda_{1}^{2}+\lambda_{2}^{2})}+\lambda\,\rho
+\lambda_{1}\,\rho_{1}+\lambda_{2}\,\rho_{2}\right).
\ee
Applying the saddle point equations
$$
\frac{\partial E_{\pm}}{\partial \lambda}=0,\;\;
\frac{\partial E_{\pm}}{\partial \lambda_{1,2}}=0,
$$
after some algebra, gives
\be
\lambda_{1,2}=\mp \frac{|\lambda|\,\rho_{1,2}}{2 \sqrt{1-\rho_{1}^{2}
-\rho_{2}^{2}}},\;\;\rho_{1}^{2}+\rho_{2}^{2}= 4 (\rho -\rho^{2}).
\ee
With the help of these equations, the dependence on $\lambda_{1,2}$
and $\rho_{1,2}$ in (54) can be eliminated, and $E_{\pm}$ can be
expressed in terms of only $\rho$;
\be
E_{\pm}= 4 \frac{p^{+}}{a} \frac{g^{+}}{a^{2}} (\rho -\rho^{2}) B(\rho). 
\ee

The above expression for the ground state energy is the basic result of this
section. We note that:\\
a) Unlike in the case of the $\phi^{3}$ theory (eq.(45)),
there is no splitting of the ground state energy; for either sign of the
square root in (54), the energy is the same.\\
b) Remembering that $B(\rho)$ is positive and bounded in the physical
range of $\rho$, the ground state energy is minimized for either
$\rho=0$ or for $\rho=1$, with vanishing energy in both cases.
 The first possibility corresponds to the
trivial situation of an empty world sheet with no Feynman graphs. The second
case is equally trivial: It corresponds to a world sheet
completely covered with eternal solid lines, propagating freely with no
interaction. So unlike in the case of the  $\phi^{3}$ theory, the ground
state is not a condensate.\\
c) In reaching this conclusion, we have taken $g^{2}$ to be positive, in
order to have a stable $\phi^{4}$ theory. If, instead, we take $g^{2}$
to be negative, the ground state energy becomes negative. Since it also
vanishes at the two end points, there is clearly a minimum at some
value of $\rho$ different from zero or one. The ground state is therefore
a condensate for the ``wrong sign'' $\phi^{4}$. It is of some interest
to note that early work on string formation in field theory [9,10] was also
based on  $\phi^{4}$ theory with negative $g^{2}$.

It is no accident that there is no condensation for the positive sign
of $g^{2}$. In this case, the interaction between the $\phi$ particles
is repulsive, and evidently
an attractive force is needed to form a condensate. This attractive
force is present in the $\phi^{3}$ model or the wrong sign $\phi^{4}$,
but unfortunately, both models are unstable. It remains to see whether
non-abelian gauge theories, despite being stable, can still support
a condensate on the world sheet.

\vskip 9pt

\noindent{\bf 8. Conclusions And Future Directions}

\vskip 9pt

The summation of the planar Feynman graphs using world sheet techniques
 continues to be one of the more interesting and challenging problems
in contemporary research [11]. The approach pursued by the present author
and C.B.Thorn [1,2,3,5] relied on the methods of field theory applied
to the world sheet. The starting point was the path integral based on
a suitable action on the world sheet; given the action, the meanfield
approximation was used to solve for the ground state of the theory.
This was the approach used in [5]; however,
a serious problem with [5] was the necessity of an ultraviolet cutoff
to get finite answers. In this article, we do not directly use the
path integral to quantize the theory; instead, starting with the classical
action, we formulate the dynamics in terms of the Hamiltonian and the
canonical commutation relations. The resulting theory is free of the
ultraviolet divergence, which is the main result of the present work.

Another advantage of the present  approach is that the world sheet
field theory can be initially formulated exactly, before resorting to
any approximations. Furthermore, in addition to the $\phi^{3}$, on which
most of the previous work was based, we are able to extend our
treatment to the $\phi^{4}$ theory.

In the later part of the paper, we show that, using the mean field
approximation, it is possible to extract some information from this
formalism. In particular, we are interested in the structure of the
ground state of the model. There are two alternative pictures for
the ground state: Either it is perturbative or the Feynman graphs
form a dense network (condensate) on the world sheet. In the case of
the $\phi^{3}$ model and the $\phi^{4}$ model with the ``wrong''
sign of the coupling constant, we find that the condensate phase is
the one that is favored energetically, whereas for the $\phi^{4}$
model with the physical sign of the coupling, the perturbative ground
state has the lower energy. For the $\phi^{3}$ theory,
 this confirms  the results of
[2,5]; the main advance over the earlier work is that the present work is 
on a more solid footing, since no ultraviolet cutoff was needed in deriving
them. 

Much still remains to be done. For example, there is an infrared
cutoff in the model in the form of the compactification of the
lightcone coordinate $x^{-}$; at least in the mean field approximation,
it should be possible to decompactify the model by removing this cutoff.
 Another interesting
project is to go beyond the leading order in the mean field method
and compute the non-leading terms. These non-leading contributions are
important, since they contain  information about the spectrum
of the model. The knowledge of the spectrum will help answer the
question whether the condensation of Feynaman graphs on the world
sheet results in the formation of a string.
 A spectrum  that consists of asymptotically linear
trajectories would signal string formation on the world sheet.
 Also, we expect the $\phi^{3}$ theory and the $\phi^{4}$ theory
with wrong sign of the coupling to be unstable;
 the knowledge of the spectrum should give some information about
 the stability of the model. Finally,
using the tools introduced in this paper, especially the technique of
projection operators, it should be possible to tackle physically more
relevant models, such as non-abelian gauge theories.

\vskip 9pt

\noindent{\bf Acknowledgement}

\vskip 9pt

This work was supported  by the Director, Office of Science,
 Office of Basic Energy Sciences, 
 of the U.S. Department of Energy under Contract DE-AC02-05CH11231.

\vskip 9pt

\noindent{\bf References}
\begin{enumerate}
\item K.Bardakci and C.B.Thorn, Nucl.Phys. {\bf B 626} (2002) 287,
hep-th/0110301.
\item K.Bardakci and C.B.Thorn, Nucl.Phys. {\bf B 652} (2003) 196,
hep-th/0206205.
\item S.Gudmundsson, C.B.Thorn and T.A.Tran, Nucl.Phys. {\bf B 649}
(2003) 3,  hep-th/0209102, C.B.Thorn and T.A.Tran, Nucl.Phys. {\bf B 677}
(2004) 289, hep-th/0307203.
\item G.'t Hooft, Nucl.Phys. {\bf B 72} (1974) 461.
\item K.Bardakci, Nucl.Phys. {\bf B 715} 141, hep-th/0501107.
\item T.Banks, W.Fischler, S.H.Shenker,and L.Susskind, Phs.Rev.
{\bf D 55} (1997) 5112, hep-th/9610043, L.Susskind, hep-th/9704080.
\item N.Seiberg, Phys.Rev.Lett. {\bf 79} (1997) 3577, hep-th/9710009.
\item S.R.Coleman and E.Weinberg, Phys.Rev. {\bf D 7} (1973) 1888.
\item H.B.Nielsen and P.Olesen, Phys.Lett. {\bf B 32} (1970) 203,
\item B.Sakita and M.A.Virasoro, Phys.Rev.Lett. {\bf 24} (1970) 1146.
\item For some alternative approaches for putting field theory
on the world sheet, see O.Aharony, J.R.David, R.Gopakumar, Z.Komargodski
and S.S.Razamat, Phys.Rev. {\bf D 75} (2007) 106006, O.Aharony,
Z.Komargodski and S.S.Razamat, JHEP {\bf 0605} (2006) 016,
hep-th/0602226,
A.Clark, A.Karch, P.Kovtun and D.Yamada, Phys.Rev. {\bf D 68} (2003)
066011, hep-th/0304107, M.Kruczenski, hep-th/0703218.
 
\end{enumerate}

\end{document}